\documentclass[prx,preprint,aps,amsmath,amssymb,showpacs]{revtex4-2}
\usepackage{graphicx}
\usepackage{caption}

\begin{document}
\preprint{}
\draft

\title{Convert laser light into single photons via interference}

\author
{Yanfeng Li$^{1,4,5}$$^{\dag}$, Manman Wang$^{1,4,5}$$^{\dag}$, Guoqi Huang$^1$, Li Liu$^1$, Wenyan Wang$^1$, Weijie Ji$^1$,
Hanqing Liu$^{2,3}$, Xiangbin Su$^{2,3}$, Shulun Li$^{2,3}$,  Deyan Dai$^{2,3}$,  Xiangjun Shang$^{2,3}$,
Haiqiao Ni$^{2,3}$, Zhichuan Niu$^{2,3}$${^\ast}$ and Chengyong Hu$^1$}
\thanks{Corresponding authors. e-mail: cyhu03@gmail.com; zcniu@semi.ac.cn\\
$\dag$ These authors contributed equally to this work.}
\affiliation{$^1$Beijing Academy of Quantum Information Sciences, Beijing 100193, China.}
\affiliation{$^2$State Key Laboratory of Superlattices and Microstructures,
Institute of Semiconductors, Chinese Academy of Sciences, Beijing 100083, China.}
\affiliation{$^3$College of Materials Science and Opto-Electronic Technology,
University of Chinese Academy of Sciences, Beijing 101408, China.}
\affiliation{$^4$Institute of Physics, Chinese Academy of Sciences, Beijing 100190, China}
\affiliation{$^5$University of Chinese Academy of Sciences, Beijing 101408, China.}

\begin{abstract}
Laser light possesses perfect coherence, but cannot be attenuated to single photons via linear optics.
An elegant route to convert laser light into single photons is based on photon blockade
in a cavity with a single atom in the strong coupling regime.
However, the single-photon purity achieved by this method remains relatively low.
Here we propose an interference-based approach where laser light can be transformed into
single photons by destructively interfering with a weak but super-bunched incoherent field
emitted from a cavity coupling to a single quantum emitter. Fully destructive interference
erases the two-photon probability amplitude in the laser field and results in anti-bunching.
We demonstrate this idea by measuring the reflected light of a laser field
which drives a double-sided optical microcavity containing a single artificial atom - quantum dot (QD) in the Purcell regime.
The reflected light consists of a superposition of the driving field with the cavity output field.
We achieve the second-order autocorrelation $\mathrm{g}^{(2)}(0)=0.030\pm 0.002$ and the two-photon
interference visibility $94.3\%\pm 0.2\%$. By separating the coherent
and incoherent fields in the reflected light, we observe that the incoherent field from the
cavity exhibits super-bunching with $\mathrm{g}^{(2)}(0)=41\pm 2$ while the coherent field remains Poissonian statistics.
By controlling the relative amplitude of coherent and incoherent fields, we verify that photon
statistics of reflected light is tuneable from perfect anti-bunching to super-bunching
in agreement with our predictions.
Our results demonstrate photon statistics of light as a quantum interference phenomenon
and that a single QD can scatter two photons simultaneously at low driving fields in contrast to
the common picture that a single two-level quantum emitter can only scatter (or absorb and emit) single photons.
This work opens the door to tailoring photon statistics of laser light
via cavity or waveguide quantum electrodynamics and interference. Inheriting the laser's ultra-long coherence time
and robust photon indistinguishability, coherent single photons generated by this approach could be a
key resource for interference-based quantum information technologies.

\end{abstract}

\date{\today}

\maketitle

\section{INTRODUCTION}
Single photon source (SPS) that emits one photon at a time is an essential component in
quantum information science and technology, including quantum communications and
networks \cite{kimble08, wehner18, xu20, lu21},
quantum computation and simulation \cite{obrien09, chang14, pelucchi22}, quantum metrology and quantum
sensing \cite{giovannetti11, pirandola18, couteau23}.
Up to now, there are two approaches to generate single photons: One is a probabilistic
approach based on spontaneous parametric down-conversion (SPDC) or spontaneous four-wave mixing (SFWM) \cite{eisamana11},
and another is a deterministic approach exploiting spontaneous emission from single quantum
emitters \cite{aharonovich16, arakawa20}, such as cold atoms, ion traps, QDs and color centers.
In recent years, significant progress has been made on QD SPS towards an ideal source with
near-unity photon purity, high indistinguishability and high
efficiency \cite{ding16, wang16, somaschi16, wang19, tomm21}.
However, the emission-type QD SPSs suffer from short first-order coherence limited by twice the
QD's radiative lifetime, and fragile two-photon interference (TPI) \cite{hong87}
due to various noises in solid-state environment \cite{kuhlmann13a}.
Single photons with long coherence time are a key resource for quantum communications \cite{xu20, xie22, zeng22},
optical quantum computation \cite{obrien09} and quantum metrology \cite{giovannetti11, pirandola18, couteau23}
which are reliant on either TPI or single-photon interference.

An alternative way to generate single photons is to transform laser light into single photons
based on either parametric nonlinearities \cite{koashi93,lu01} or single-photon quantum
nonlinearity \cite{birnbaum05, faraon08, muller15}. Such singe photons are believed to inherit
laser's first-order coherence and photon indistinguishability.
Based on the photon blockade effect \cite{tian92, imamoglu97}
where a first photon from the incident laser blocks the transmission of a second photon
due to the anharmonic spacings in the Jaynes-Cummings energy ladder,
Birnbaum et al \cite{birnbaum05} converted an
incident laser light into single photons with $\mathrm{g}^{(2)}(0)=0.13$ in an optical
cavity with one trapped atom in the strong coupling regime.
Later, Faraon et al \cite{faraon08} and M\"{u}ller et al \cite{muller15} reported
similar results with $\mathrm{g}^{(2)}(0)=0.29$ in a photonic crystal nanocavity containing
a single QD. Further improving
the single-photon purity is possible using homodyne or self-homodyne technique \cite{muller16, fischer17}.

In this work, we propose an interference-based approach and demonstrate it in a laser-converted
single photon source (LCSPS) with ultra-long coherence time and robust photon indistinguishability
using a single QD in a double-sided optical microcavity in the Purcell regime.
This approach is similar to previous work using SPDC photon pairs \cite{lu01}. We directly measure
the reflected light of an incident laser driving the cavity and achieve the second-order autocorrelation
$\mathrm{g}^{(2)}(0)=0.030\pm 0.002$ and the degree of photon indistinguishability $94.3\%\pm 0.2\%$.
We demonstrate that a single QD scatters two photons simultaneously at low driving fields and
unambiguously affirm the interference picture on anti-bunching in resonance fluorescence proposed 40 years ago \cite{dalibard83}.
This work opens the door to tailoring photon statistics of laser light
via cavity or waveguide quantum electrodynamics (QED) \cite{raimond01, reiserer15, lodahl15, chang18}
and interference \cite{carreno18, casalengua20, hanschke20, phillips20, masters23, cordier23}.

\section{Working principle of LCSPS}
We consider a two-level quantum emitter such as QD which is resonantly coupled to a symmetric
double-sided cavity [see Fig. 1(a)] in the Purcell regime with cooperativity parameter
$C=2g^2/(\kappa \gamma_{\perp}) \gg 1$ and critical photon number
$n_0=\gamma_{\perp}\gamma_{\parallel}/(4g^2) \ll 1$, where $g$ is the QD-cavity interaction strength,
$\kappa$ is the cavity photon decay rate, $\gamma_{\parallel}$ is the QD spontaneous emission rate into leaky modes,
$\gamma_{\perp}=\gamma_{\parallel}/2+\gamma^*$ is the QD polarization decay rate, and
$\gamma^*$ is the QD pure dephasing rate.
Such design allows the driving laser field interacts with the QD deterministically,
cavity-enhanced coherent scattering and strong nonlinearity at the single-QD and single-photon
level. The cooperativity parameter describes the QD polarization decay rate into the cavity
mode is C times that into the leaky modes.
The larger is the cooperativity parameter, the higher is the QD-induced coherent reflectivity
$R_{QD}=[C/(1+C)]^2$ and the lower is the QD-induced coherent transmissivity $T_{QD}=1/(1+C)^2$
[see Appendix A, Eq. (\ref{eqa3})]. The critical photon number describes the number of
cavity photons required to saturate the QD response. For $n_0 < 1$, a single
photon inside the cavity would induce significant changes in cavity properties, which is
the single-photon nonlinearity.

If the QD decouples to the cavity (e.g., no QD in the cavity or QD gets saturated),
the cavity (we call it cold cavity thereafter) is transmissive at the
central frequency of cavity mode [Figs. 1(c) and 1(e), blue curves].
If the QD couples to the cavity, the cavity (we call it hot cavity thereafter)
is reflective at the central frequency of cavity mode with a reflection
peak at the QD resonance [Figs. 1(c) and 1(e), red curve]. This is the
single-QD nonlinearity.

The cavity is resonantly driven by a laser field with the amplitude $a_{in}$ [see Fig. 1(a)].
The transmitted field which is the cavity-emitted field $\sqrt{\kappa}\hat{a}$ is admixed with a local laser
field with the amplitude $E_{LO}$
on a 50:50 non-polarizing beam splitter. The local oscillator is set
in phase or $180^{\circ}$ out of phase with the driving field.
One of the outputs of this beam splitter contains the superimposed light (SL) with
the field operator
\begin{equation}
\hat{a}_{SL}=E_{LO}+\sqrt{\kappa}\hat{a},
\label{eq1}
\end{equation}
where $\hat{a}$ is the cavity field operator and $\kappa/2$ is the cavity field decay rate into
the reflection or transmission channels.
The SL field can be re-written as $\hat{a}_{SL}=\langle \hat{a}_{SL}\rangle+\delta \hat{a}_{SL}$
where $\langle \hat{a}_{SL}\rangle=E_{LO}+\sqrt{\kappa}\langle\hat{a}\rangle$ is
the coherent component with $\delta$-like laser spectrum
and $\delta \hat{a}_{SL}=\sqrt{\kappa}\delta\hat{a}$ describes the incoherent component
due to quantum fluctuations with broad QD emission spectrum. There is no mutual first-order coherence
between coherent and incoherent components as $\langle \delta\hat{a}\rangle=0$.
The dynamics and photon statistics of the cavity and SL fields
can be calculated using the master equation (see Appendix A).

Fig. 1(b) presents $\mathrm{g}^{(2)}(0)$ of SL as a function of the relative
amplitude of the local field $E_{LO}$ to the driving field $a_{in}$.
When $E_{LO}=0$, the SL field which is exactly
the transmitted field $\hat{a}_{t}=\sqrt{\kappa}\hat{a}$ exhibits strong
super-bunching with $\mathrm{g}^{(2)}(0)\simeq182$ [also see Fig. 1(d)]. When $E_{LO}=a_{in}$, the SL field which is exactly
the reflected field $\hat{a}_{r}=a_{in}+\sqrt{\kappa}\hat{a}$ exhibits strong
anti-bunching with $\mathrm{g}^{(2)}(0)=0.005$. At $E_{LO}=a_{in}/(1+C)$,
the local laser field cancels the coherent cavity-output field $\sqrt{\kappa}\langle \hat{a} \rangle=-a_{in}/(1+C)$
[Note that the transmitted field and the cavity output field
are $180^{\circ}$ out of phase with the driving field, see Eq. (\ref{eqa2}) in Appendix A].
The SL field which contains the pure incoherent cavity output exhibits the strongest super-bunching with $\mathrm{g}^{(2)}(0)\simeq 600$.
When $E_{LO}=-a_{in}$, the SL field exhibits weak anti-bunching with $\mathrm{g}^{(2)}(0)=0.47$.
When $E_{LO}\gg a_{in}$ or $E_{LO} \ll -a_{in}$, the SL field shows Poissonian statistics with $\mathrm{g}^{(2)}(0)=1$
as the driving laser.
These results demonstrate the photon statistics of SL as an interference phenomenon
which is tunable from perfect anti-bunching to super-bunching by controlling
the relative amplitude of the local field to the driving field.

Due to single-photon nonlinearity, the injection of one photon into the cavity
turns the QD-cavity system from reflective to transmissive [see Figs. 1(c) and 1(e)], so a second
photon would pass through the cavity. This phenomenon is called photon-induced tunneling (PIT) \cite{faraon08}.
Thus the cavity output field into the transmission and reflection channels [see Fig. 1(a)] contains
two-photon and multi-photon bound states \cite{shen07}
and exhibits bunching or super-bunching [see Fig. 1(b)]. The two-photon process resembles the degenerate SFWM where
two laser photons are scattered into idler and signal photons.
At low driving fields, we could neglect the multi-photon
components ($n \geq 3$) and describe the incoherent cavity output field as a two-photon state
$|\psi\rangle_{cav}^{inc}=|0\rangle-(\xi/\sqrt{2})|2\rangle$ where the negative sign
accounts for the $180^{\circ}$ phase shift between cavity output field and the driving field.
$\xi$ is determined by the intensity of incoherent component and is related to the driving field by
 $\xi^2 = (a_{in}^2/\Gamma_{\parallel})^2$
[see Appendix B, Eq. (\ref{eqb4})], where $\Gamma_{\parallel}$ is the
cavity-enhanced QD emission rate, i.e., $\Gamma_{\parallel}=(F_P+1)\gamma_{\parallel}$ with
$F_P$ the Purcell factor and $\gamma_{\parallel}$  the QD spontaneous emission rate into leaky modes.
The local field in phase with the driving field is described by a coherent state
$|\alpha\rangle=|0\rangle+\alpha|1\rangle+(\alpha^2/\sqrt{2})|2\rangle$ for $|\alpha|\ll 1$ with $\alpha$
related to the field amplitude by $\alpha^2 = E_{LO}^2/\Gamma_{\parallel}$.
The two-photon probability $(\alpha^2-\xi) /\sqrt{2}$ in the SL field
$|\alpha\rangle+|\psi\rangle_{cav}^{inc}$
disappears at $E_{LO}=a_{in}$, so SL exhibits anti-bunching as shown in Fig. 1(b).
Destructive interference between a coherent state and a two-photon state generated
by SPDC was reported in previous work \cite{lu01}. Note that the two-photon scattering process
also occurs at high driving fields where incoherent components dominate and Mollow triplet is
observed \cite{ulhaq12}.

To demonstrate this approach for LCSPS at $E_{LO}=a_{in}$, we investigate the photon statistics of reflected
light from the QD-cavity system in more details. The reflected light is a superposition of the driving field
and the cavity output field.
The calculated $\mathrm{g}^{(2)}(0)$ of reflected light exhibits laser frequency detuning
dependence [Fig. 1(f)] and laser power dependence [Fig. 1(g)]. These dependence
are both related to cavity and can be well explained by the interference picture discussed above.
The frequency detuning causes a decrease of the QD-induced reflectivity $R_{QD}$ [see Fig. 1(e)],
the increased fraction of incoherent field leads to a rise of $\mathrm{g}^{(2)}(0)$ [see Fig. 1(f)].
At the two reflectivity minima ($R_{QD}=0$) [see Fig. 1(e)], the reflected light contains only the incoherent
field and exhibits the strongest super-bunching [see Fig. 1(f)].
With increasing the driving field, the increased fraction of incoherent field (see Appendix B)
also induces a rise of $\mathrm{g}^{(2)}(0)$ [see Fig. 1(g)].
We are interested in the strong anti-bunching with $\mathrm{g}^{(2)}(0)=10^{-4}-10^{-3}$
at lower driving fields with $\Omega/\Gamma_{\parallel} < 1/\sqrt{2}$.
In this work we adopt Rabi frequency $\Omega \propto a_{in} \propto \sqrt{P_{laser}}$
to describe the driving laser power and normalize it to the cavity-enhanced QD emission rate $\Gamma_{\parallel}$.

\section{Implementation of LCSPS}
We fabricated the pillar cavity [see inset of Fig. 1(g)]
containing a single QD resonantly coupling to the fundamental cavity mode with the cooperativity parameter $C=6.9$ and
the critical photon number $n_0=6.9\times 10^{-4}$. The details for sample growth and device fabrication can be found in
Appendix C. The two distributed Bragg reflectors (DBRs) are asymmetric in the realistic devices such that the leakage rate
from the top mirror can balance the sum of the leakage rate from the bottom mirror, cavity side leakage and
background absorption resided in the mirror and cavity materials.
Such cavity design allows nearly zero reflectivity ($R=0.89\%$) at the center of
the fundamental cavity mode [Fig. 2(c), blue dots].
The QD transition induces a reflectivity peak with $R=46.6\%$ inside a broad cavity resonance [Fig. 2(c), red dots].
The measured reflection spectra for cold and hot cavities agree well with calculated results
[see Fig. 1(e)] using Eq. (\ref{eqa2}) in Appendix A except that the measured QD reflectivity $R_{QD}$ is lower than
theoretical value [see Fig. 2(b)].
This could arise from the QD blinking effect which reduces the QD-cavity interaction strength $g$.
With increasing the driving power, the QD reflectivity peak tends to wash out for $\Omega > \Gamma_{\parallel}$
due to the QD saturation [Figs. 2(a) and 2(b)].
The slight increase in the QD reflectivity [Fig. 2(b), blue dots] in lower power region ($\Omega < 0.1\Gamma_{\parallel}$)
is unexpected and needs further investigation.

Fixing the laser frequency on the QD resonance, we measure
the second-order autocorrelation function of the reflected light [Fig. 2(d)] and
get $\mathrm{g}^{(2)}(0)=0.030\pm 0.002$ (raw data), indicating good single photon nature.
The experimental $\mathrm{g}^{(2)}(\tau)$ curve [Fig. 2(d), red solid line) agrees well with the
calculations [Fig. 2(d), blue dotted line] using the master equation in Appendix A
or Eq. (\ref{eqb7}) in Appendix B.
Taking the QD radiative lifetime of $57$ ps (see Appendix C, Fig. 8),
the fitting with Eq. (\ref{eqb7}) yields  $\gamma^{*}= 0.03\Gamma_{\parallel}$, indicating
the pure dephasing is negligible at low driving fields.
$\mathrm{g}^{(2)}(0)$ shows the laser-detuning dependence [see Fig. 2(e), blue dots]
and the laser-power dependence [see Fig. 2(b), red dots],
in agreement with the theoretical predictions [see Figs. 1(f) and 1(g)].
Note that $\mathrm{g}^{(2)}(0)$ remains the lowest level around $0.030$ at low driving fields with
$\Omega < \Gamma_{\parallel}/\sqrt{2}$ as predicted in Fig. 1(g).

Now we analyze the photon statistics of coherent and incoherent components in the reflected light.
The coherent cavity output field inherits the laser spectrum of driving field while the
incoherent cavity output field due to quantum fluctuations has the broad QD emission spectrum.
As the driving laser linewidth ($ < 100\ $kHz)
is much smaller than the cavity-enhance QD linewidth $\Gamma_{\parallel}=2.8\ $GHz (see Appendix C, Fig. 8),
we could use a Fabry-Perot filter with a resolution of $31\ $MHz to separate coherent
component from incoherent components [see Fig. 3(a)] and measure their photon statistics separately.

It is not surprising that the coherent components passing through the Fabry-Perot filter
show Poissonian statistics [see Fig. 3(b)]. However, the remaining field which is reflected from the Fabry-Perot filter
changes photon statistics from anti-bunching with $\mathrm{g}^{(2)}(0)=0.030\pm 0.002$ to super-bunching with $\mathrm{g}^{(2)}(0)=41\pm 2$ with increasing the filter's transmission from $T_F=0$  to $T_F=63\%$
via frequency detuning [see Fig. 3(c)].
As the coherent cavity output field $\langle \hat{a} \rangle = -a_{in}/[(1+C)]$ at $\Delta=0$ is much
weaker than the driving field amplitude $a_{in}$,
we could simulate the filtering effect by modifying the reflected field operator as
$\hat{a}_{F}=\sqrt{1-T_F}a_{in}+\sqrt{\kappa}\hat{a}$, where $T_F=0$ corresponds to the reflected field
operator $\hat{a}_{r}$ and $T_F=1$ corresponds to the transmitted field operator $\hat{a}_{t}$.
The calculated results in Fig. 3(d) using the master equation (see Appendix A) agree well with experimental curves [see Fig. 3(c)].
To our knowledge, it is the first time to observe that the photon statistics changes from perfect antibunching
to super-bunching when rejecting one of the scattering components \cite{hanschke20, phillips20, masters23, cordier23}.
The measured super-bunching is smaller than theoretical value due to the incomplete rejection of
the coherent components and the finite timing resolution ($\sim 20\ $ps) of the correlation measurement kit.

The super-bunching of the cavity output field (or the transmitted light) demonstrate
that a single two-level quantum emitter can scatter two photon simultaneously proposed in previous work \cite{dalibard83, shen07}.
Our results show photon statistics of reflected and transmitted light
is an interference phenomenon.  The driving field is converted to single photons
by destructively interfering with a weak incoherent field emitted from the cavity.
Our experiments affirm the interference picture on the origin of anti-bunching in
resonance fluorescence \cite{dalibard83, masters23}.

As we directly collect the reflected light without using the cross-polarization configuration \cite{kuhlmann13b}
which is often used in the emission-type SPSs to suppress the laser scattering down to $10^{-7}-10^{-5}$ levels,
LCSPS could achieve high single-photon count rate and high laser conversion efficiency.
The single-photon count rate detected with a superconducting nanowire single-photon detector (SNSPD)
increases linearly with the driving laser power up to $\Omega=\Gamma_{\parallel}/\sqrt{2}$ [see Fig. 2(f)].
A maximum single-photon count rate of $\sim 200$ MHz is achieved with an overall conversion efficiency
of $\eta \sim 2\%$ for this device, where $\eta$ is defined as the ratio of the detected single photon rate to
the driving-laser photon rate.
The conversion efficiency of LCSPS is several orders of magnitude higher than $\eta \sim 10^{-10}-10^{-6}$
of SPDC heralded sources \cite{tanzilli01}, and $\eta \sim 10^{-4}$ of the emission-type
QD SPSs \cite{ding16, wang16, somaschi16, wang19, tomm21}. As long as the QD-induced coherent reflectivity
$R_{QD}\rightarrow 1$, the driving laser light could be completely transformed into single photons with
the theoretical conversion efficiency of $100\%$.  As our confocal microscopy does not collect all the reflected light
and there are various losses from optical components in the detection path (see Appendix C, Fig. 7),
the practical conversion efficiency $\sim 2\%$ is much lower than the theoretical value.

\section{First-order coherence}
From the reflection field operator $\hat{a}_{r}=a_{in}+\sqrt{\kappa}\hat{a}$, we get the degree of the first-order coherence
\begin{equation}
\mathrm{g}^{(1)}(\tau)=\frac{I_{coh}}{I_{tot}}\mathrm{g}_L^{(1)}(\tau)+\frac{I_{incoh}}{I_{tot}}\mathrm{g}_{inc}^{(1)}(\tau),
\label{eq4}
\end{equation}
where $I_{coh}$ and $I_{incoh}$ are the intensity of coherent and incoherent components of reflected light, respectively.
As there is no mutual coherence between coherent and incoherent components, the total intensity is
$I_{tot}=I_{coh}+I_{incoh}$. The coherent component shows laser's coherence $\mathrm{g}_L^{(1)}(\tau)$ and
$\delta$-like laser spectrum, while the incoherent component has short coherence time determined by the QD
radiative lifetime and pure dephasing time. An analytical expression for $\mathrm{g}^{(1)}(\tau)$ is derived using optical Bloch equations
[see Appendix B, Eq. (\ref{eqb5})].

Fig. 4(a) presents the measured degree of first-order coherence $\mathrm{g}^{(1)}(\tau)$ of reflected light versus
time delay with a Michelson interferometer. At higher driving fields, there are two coherence times identified,
$\tau_{c1} \simeq 115$ ps and  $\tau_{c2}> 24.5$ ns which is limited by the longest path delay of interferometer.
The short 115-ps coherence time which is twice the QD radiative lifetime $57\ $ps (see Appendix C, Fig. 8)
comes from the incoherent field emitted from the cavity, while the long coherence time comes from
the driving laser with ultra-long coherence time ($>10\ \mu$s).
Single photons with long coherence time were also reported in previous work \cite{nguyen11, matthiesen12}.
The incoherent components reduce the degree of
first-order coherence by the intensity fraction of total incoherent components.
The ratio of coherent to incoherent components decreases with increasing the driving power, so does the
degree of first-order coherence [see Figs. 4(a) and 4(b), dots]. The experimental results
agree well with theoretical curves [see Figs. 4(a) and 4(b), solid lines]
using either master equation in Appendix A or Eq. (\ref{eqb5}) in Appendix B.

At low driving fields with $\Omega < \Gamma_{\parallel}/\sqrt{2}$ where the incoherent cavity-emitted field
is much weaker than the driving field, $|\mathrm{g}^{(1)}(\tau)| \geq 0.5$ for all $\tau$ over the
driving laser's coherence time ($>10\ \mu$s), in this sense we say the LCSPS inherits the laser's first-order coherence time.
The increase in $|\mathrm{g}^{(1)}(\tau)|$ at strong driving fields ($\Omega > \Gamma_{\parallel}$)
[see Fig. 4(b), dots] is due to the strong laser background when QD saturates.
The oscillations in the calculated curves [Fig. 4(b), solid lines] may come from the interference
between three incoherent components [see Appendix B, Eq. (\ref{eqb5})].

\section{Photon indistinguishability}

To characterize the photon indistinguishability of LCSPS, we performed Hong-Ou-Mandel(HOM) interference
experiment using an asymmetric Mach-Zehnder interferometer(AMZI) [see Fig. 5(a)].
As LCSPS inherits the driving laser's first-order coherence ($>10\ \mu$s) which is much
longer than the AMZI path delays, classical interference which is neglected in
previous work \cite{patel08, proux15} needs to be taken into account here.
The modified expressions for $\mathrm{g}^{(2)}_{\perp}(\tau)$ and $\mathrm{g}^{(2)}_{\parallel}(\tau)$
in cross- and parallel-polarization configurations are
\begin{equation}
\begin{split}
\mathrm{g}^{(2)}_{\perp}(\tau)=&\frac{1}{N}\Biggl\{(R_A^2+T_A^2)R_BT_B\mathrm{g}^{(2)}(\tau)\\
&+R_B^2R_AT_A\mathrm{g}^{(2)}(\tau+\Delta t)+T_B^2R_AT_A\mathrm{g}^{(2)}(\tau-\Delta t)\Biggl\},\\
\mathrm{g}^{(2)}_{\parallel}(\tau)=&
\frac{1}{N}\Biggl\{(R_A^2+T_A^2)R_BT_B\mathrm{g}^{(2)}(\tau)\\
&+R_B^2R_AT_A\mathrm{g}^{(2)}(\tau+\Delta t)+T_B^2R_AT_A\mathrm{g}^{(2)}(\tau-\Delta t)\\
&-2R_AT_AR_BT_B V_0|\mathrm{g}^{(1)}(\tau)|^2\Biggl\},
\end{split}
\label{eq3}
\end{equation}
where $N=(R_A^2+T_A^2)R_BT_B+(R_B^2+T_B^2)R_AT_A$, $R_{A,B}$ and $T_{A,B}$ are the reflection and transmission intensity
coefficients of two beam splitters with $R_A=T_A=50\%$ and $R_B=T_B=50\%$, and the parameter $V_0$ accounts
for the wave packet overlap on the second beam splitter $\mathrm{BS_B}$.
$\mathrm{g}^{(1)}(\tau)$ and $\mathrm{g}^{(2)}(\tau)$ are the first- and second-order correlation functions
of reflected light. The HOM (or TPI) visibility is defined as
$V_{HOM}(\tau)=[\mathrm{g}^{(2)}_{\perp}(\tau)-\mathrm{g}^{(2)}_{\parallel}(\tau)]/\mathrm{g}^{(2)}_{\perp}(\tau)$.
It is well known that classical interference leads to a broad HOM dip with a depth of $0.5$ and a width of coherence time
in an AMZI if the light intensities are equal in two arms \cite{paul86, rarity05}, therefore we
normalize $\mathrm{g}^{(2)}_{\parallel}(\tau)$ to $0.5$ in our work,
rather than $1$ in previous work \cite{patel08, proux15}.

In cross-polarization configuration [Figs. 5(b) and 5(d), blue curves], both classical interference and TPI
are not expected and we observe three $\mathrm{g}^{(2)}_{\perp}(\tau)$ dips at $\tau=0, \pm \Delta t$
due to the single-photon nature of LCSPS (two side peaks are not shown for $\Delta t=5\ \mu$s as
the time delay goes beyond our measurement range).

In parallel-polarization configuration [Figs. 5(b) and 5(d), red curves], both TPI and the
photon statistics contribute to the anti-bunching dip at $\tau=0$ and the
classical interference leads to a 0.5 background in $\mathrm{g}^{(2)}_{\parallel}(\tau)$
on the time scale of laser's coherence time ($>10\ \mu$s). The bunching peaks at $\tau=\pm \Delta t$
are due to the classical interference and intricate photon correlations.
However, this phenomenon does not affect the measured values of photon indistinguishability
and we leave detailed discussions elsewhere \cite{wang24}.

We measure the TPI visibility $V_{HOM}(0)=94.3\%\pm 0.2\%$  at $\Delta t=2$ns [Fig. 5(c)]
and $V_{HOM}(0)=93.7\%\pm 0.2\%$ at $\Delta t=5\ \mu$s [Fig. 5(e)].
Note that these are raw data without correcting experimental imperfections such as limited
detector resolution, non-unity photon purity, and imperfect mode overlap in the beam splitter.

At lower driving powers $\Omega < \Gamma_{\parallel}/\sqrt{2}$ where photons from the driving laser dominate
the reflected light, LCSPS inherits the laser's photon indistinguishability which is robust
against various noises in QDs \cite{kuhlmann13a}. Therefore, there is nearly no degradation
in TPI visibility for two photons separated by $8\ $km fibre or even
longer (corresponding to the time separation $>40\ \mu$s) [Fig. 5(f)]. This means
LCSPS is suitable for scale up.

\section{DISCUSSION AND OUTLOOK}
We have demonstrated that a single QD scatters two photons simultaneously at low driving fields in contrast to
the traditional picture that a single quantum emitter can only scatter (or absorb and emit) single photons
in the context of resonance fluorescence.
The two-photon scattering process resembles degenerate SFWM and is manifested as super-bunched incoherent
component contained in both reflected and transmitted light, which controls the photon statistics.

Based on the above two-photon process, we propose and demonstrate an interference-based approach to convert
laser light into single photons by interfering with the weak, but super-bunched incoherent field emitted
from a double-sided optical cavity containing a single QD in the Purcell regime.
Fully destructive interference erases the two-photon probability in reflected light and leads to anti-bunching
at low driving fields. The two-photon correlation time of the incoherent component
determines the correlation time of converted single photons.
Our results unambiguously affirm the interference picture on anti-bunching in resonance
fluorescence proposed four decades ago \cite{dalibard83}.
Compared with the emission-type SPSs, LCSPS inherits the incident laser's first-order coherence
and photon indistinguishability which is robust against noises in the
QD environment. We believe coherent single photons generated by this approach
would find wide applications in interference-based quantum communications,
distributed quantum computing and quantum metrology.

We also demonstrate photon statistics of light as an interference phenomenon.
Although there is no mutual first-order coherence between coherent and incoherent components,
the fourth-order interference still occurs and affects the photon statistics.
By controlling the relative amplitude between the driving field and
the cavity output field, we observe that photon statistics of reflected light is tuneable
from perfect anti-bunching to super-bunching in agreement with our numerical simulations.
This work paves the way to on-demand tailoring of photon statistics of laser light via
cavity (or waveguide) QED and interference for applications in both photonic quantum technologies and
fundamental quantum optics research, such as multi-photon scattering by a single quantum emitter,
four-wave mixing in the quantum limit\cite{li24}, squeezed light with different photon statistics,
many-body physics in photonic systems, and new approaches to generate entanglement.

\begin{acknowledgments}
The authors thank X.-J. Wang and B. Wu for assisting in lab building.  This work is supported 
by the Beijing Natural Science Foundation under the grant IS23069. Z.C. Niu is grateful 
to the National Key Technology R$\&$D Program of China under the grant 2018YFA0306101 
for financial support.
\end{acknowledgments}

\appendix

\section{METHODS TO CALCULATE DYNAMICS AND COHERENT REFLECTION/TRANSMISSION SPECTRA}

We use Heisenberg-Langevin equations of motions and the master equation to calculate the
coherent reflection/transmission spectra and investigate the dynamics of cavity field and
atomic dipole polarization in QD-cavity systems.

The Heisenberg equations of motions \cite{walls94} for the cavity field operator $\hat{a}$ and the QD dipole
operators $\sigma_-$, $\sigma_z$, together with the input-output
relation \cite{gardiner85} can be written as
\begin{equation}
\begin{cases}
& \frac{d\hat{a}}{dt}=-\left[i(\omega_c-\omega)+\frac{\kappa_1}{2}+\frac{\kappa_2}{2}+\frac{\kappa_s}{2}\right]\hat{a}-g\sigma_-
-\sqrt{\kappa_1}\hat{a}_{in}-\sqrt{\kappa_2}\hat{a}^{\prime}_{in} \\
& \frac{d\sigma_-}{dt}=-\left[i(\omega_{QD}-\omega)+\gamma_{\perp}\right]\sigma_--g\sigma_z\hat{a} \\
& \frac{d\sigma_z}{dt}=2g(\sigma_+\hat{a}+\hat{a}^{\dag}\sigma_-)-\gamma_{\parallel}(1+\sigma_z) \\
& \hat{a}_{out}=\hat{a}_{in}+\sqrt{\kappa_1}\hat{a} \\
& \hat{a}^{\prime}_{out}=\hat{a}^{\prime}_{in}+\sqrt{\kappa_2}\hat{a},
\end{cases}
\label{eqa1}
\end{equation}
where the parameters $g, \gamma_{\perp}, \gamma_{\parallel}, \gamma^{*}$ are defined in the same way as
in the main text. $\omega$, $\omega_c$, $\omega_{QD}$ are the frequencies of incident laser, cavity mode,
and the QD transition respectively and we consider the resonant case with $\omega_{QD}=\omega_c=\omega_0$
in this work. $\hat{a}_{in}$ and $\hat{a}^{\prime}_{in}$ are incident field amplitudes from two end
mirrors ($\hat{a}^{\prime}_{in}=0$ is taken here).
$\kappa_1$ and $\kappa_2$ are the cavity photon decay rates into
the reflection and transmission channels, respectively and $\kappa_s$ is the cavity photon decay rate into leaky modes.
The total cavity photon decay rate is $\kappa=\kappa_1+\kappa_2+\kappa_s$. In this work, we have
$\kappa_1 \simeq \kappa_2+\kappa_s$ such that $\kappa_1 \simeq \kappa/2$.

If the correlations between the cavity field and the QD dipole are neglected (this is
called the semiclassical approximation)\cite{allen87, armen06},
$\langle \sigma_{\pm}\hat{a}\rangle =\langle \sigma_{\pm}\rangle\langle \hat{a} \rangle$
and $\langle \sigma_z\hat{a}\rangle =\langle \sigma_z\rangle\langle \hat{a} \rangle$.
The reflection and transmission coefficients in steady state can be derived as \cite{hu15}
\begin{equation}
\begin{split}
& r(\omega)=1+\frac{-\kappa_1[i(\omega_{QD}-\omega)+\gamma_{\perp}]}{[i(\omega_{QD}-\omega)+
\gamma_{\perp}][i(\omega_c-\omega)+\frac{\kappa_1}{2}+\frac{\kappa_2}{2}
+\frac{\kappa_s}{2}]-g^2 \langle \sigma_z\rangle}, \\
& t(\omega)=\frac{-\sqrt{\kappa_1\kappa_2}[i(\omega_{QD}-\omega)+\gamma_{\perp}]}{[i(\omega_{QD}-\omega)+
\gamma_{\perp}][i(\omega_c-\omega)+\frac{\kappa_1}{2}+\frac{\kappa_2}{2}
+\frac{\kappa_s}{2}]-g^2 \langle \sigma_z\rangle},
\end{split}
\label{eqa2}
\end{equation}
where $\langle\sigma_z\rangle$ is the population difference between excited state and
ground state of QD. If $\langle\sigma_z\rangle=-1$, the QD stays in the ground state
with the excited state unoccupied;
if $\langle\sigma_z\rangle=0$, QD is fully saturated, i.e., $50\%$ probability
in the ground state and $50\%$ probability in the excited state.

At low driving fields, we take $\langle \sigma_z\rangle\simeq -1$.  At zero-detuning $\Delta=\omega-\omega_0=0$,
Eq. (\ref{eqa2}) can be simplified as
\begin{equation}
\begin{split}
& r(\Delta=0)=\frac{C}{1+C}, \\
& t(\Delta=0)=-\frac{1}{1+C},
\end{split}
\label{eqa3}
\end{equation}
where $C=2g^2/(\kappa \gamma_{\perp})$ is the cooperativity parameter.
For $C\gg 1$, $r(0)\simeq 1$ and $t(0)\simeq 0$.

The reflection and transmission coefficients can be also calculated numerically by solving the master equation in
the Lindblad form\cite{walls94} using a quantum optics toolbox \cite{johansson13}.
The master equation for the QD-cavity system can be written as
\begin{equation}
\begin{split}
\frac{d\rho}{dt}= &-i[H_{JC},\rho]+(\kappa_1+\kappa_2+\kappa_s)(\hat{a}\rho \hat{a}^{\dag}-\frac{1}{2}\hat{a}^{\dag}\hat{a}\rho-\frac{1}{2}\rho\hat{a}^{\dag}\hat{a})\\
& +\gamma_{\parallel}(\hat{\sigma}_-\rho \hat{\sigma}_+ - \frac{1}{2}\hat{\sigma}_+\hat{\sigma}_-\rho-\frac{1}{2}\rho\hat{\sigma}_+\hat{\sigma}_-)+
\frac{\gamma^*}{2}(\hat{\sigma}_z\rho\hat{\sigma}_z-\rho)\\
\equiv & \mathcal{L}\rho,
\end{split}
\label{eqa4}
\end{equation}
where $\mathcal{L}$ is the Liouvillian and $H_{JC}$ is the Jaynes-Cummings Hamiltonian
with a laser field driving the cavity. In the rotating frame at the frequency of the
driving field, $H_{JC}$ can be written as
\begin{equation}
\begin{split}
H_{JC}= &(\omega_c-\omega)\hat{a}^{\dag}\hat{a}+(\omega_{QD}-\omega)\sigma_+\sigma_-\\
& +ig(\sigma_+\hat{a}-\hat{a}^{\dag}\sigma_-)+i\sqrt{\kappa_1}\hat{a}_{in}(\hat{a}-\hat{a}^{\dag}),
\end{split}
\label{eqa5}
\end{equation}

Although an analytical solution to the master equation in Eq. (\ref{eqa4}) is very difficult,
the quantum optics toolbox \cite{johansson13} provides the numerical solution to the density
matrix $\rho(t)$, with which the dynamics of cavity or atomic fields including $g^{(1)}(\tau)$,
emission spectra and $g^{(2)}(\tau)$ can be calculated as shown in Figs.1-4 and Fig. 6.

By taking the operator average in the input-output relation,
the reflection and transmission coefficients in the steady state can be calculated
using the following expression \cite{hu15}
\begin{equation}
\begin{split}
& r(\omega)=1+\sqrt{\kappa_1}\frac{\mathrm{Tr}(\rho \hat{a})}{\langle\hat{a}_{in}\rangle},\\
& t(\omega)=\sqrt{\kappa_2}\frac{\mathrm{Tr}(\rho \hat{a})}{\langle\hat{a}_{in}\rangle}.
\end{split}
\label{eqa6}
\end{equation}

\section{DERIVED FORMULAS FOR $g^{(1)}(\tau)$, EMISSION SPECTRA AND $g^{(2)}(\tau)$ OF REFLECTED LIGHT}

When the incident laser is resonant with the QD transition, i.e., $\omega_L=\omega_{QD}$, we could get analytical results on
the dynamics if taking the QD-cavity system as an entirety. We extend the optical Bloch equations in text books \cite{loudon03}
to include the pure dephasing rate $\gamma^{*}$ of the dressed QD driven by a laser field at a Rabi frequency $\Omega$
\begin{equation}
\begin{cases}
& \dot{\rho}_{gg}=i\frac{\Omega}{2}(\rho_{ge}-\rho_{eg})+\Gamma_{\parallel}\rho_{ee} \\
& \dot{\rho}_{ee}=-i\frac{\Omega}{2}(\rho_{ge}-\rho_{eg})-\Gamma_{\parallel}\rho_{ee} \\
& \dot{\rho}_{ge}=[i(\omega_{QD}-\omega_L)-(\frac{\Gamma_{\parallel}}{2}+\gamma^{*})]\rho_{ge}-i\frac{\Omega}{2}(\rho_{ee}-\rho_{gg})\\
& \dot{\rho}_{eg}=[-i(\omega_{QD}-\omega_L)-(\frac{\Gamma_{\parallel}}{2}+\gamma^{*})]\rho_{eg}+i\frac{\Omega}{2}(\rho_{ee}-\rho_{gg}),
\end{cases}
\label{eqb1}
\end{equation}
where $\omega_L$, $\omega_{QD}$ are the frequencies of incident laser
and the QD transition, respectively. $\Gamma_{\parallel}=(F_P+1)\gamma_{\parallel}$ is
the cavity-enhanced QD spontaneous emission rate  and $\gamma^*$ is the QD pure
dephasing rate.

Solving the above equations, we obtain the steady-state coherent and incoherent scattering rates
\begin{equation}
\begin{split}
&I_{coh}=\Gamma_{\parallel}\left[\frac{\Gamma_{\parallel}}{(\Gamma_{\parallel}+2\gamma^{*})}\frac{S}{(1+S)^2}\right], \\
&I_{incoh}=\Gamma_{\parallel}\left[\frac{\Gamma_{\parallel}S+2\gamma^*(1+S)}{(\Gamma_{\parallel}+2\gamma^{*})}\frac{S}{(1+S)^2}\right],
\end{split}
\label{eqb2}
\end{equation}
where S is the saturation parameter
\begin{equation}
 S=\frac{\Omega^2\left(\frac{1}{2}+\frac{\gamma^{*}}{\Gamma_{\parallel}}\right)}{(\omega_{QD}-\omega_L)^2+\left(\frac{\Gamma_{\parallel}}{2}+\gamma^{*}\right)^2}.
 \label{eqb3}
\end{equation}

At $\omega_L=\omega_{QD}$ and $\gamma^{*}=0$,
Eqs. (\ref{eqb2}) and (\ref{eqb3}) reduce to
\begin{equation}
\begin{split}
&I_{coh}=\Gamma_{\parallel}\left[\frac{S}{(1+S)^2}\right], \\
&I_{incoh}=\Gamma_{\parallel}\left[\frac{S^2}{(1+S)^2}\right]=SI_{coh},\\
&S=2\left(\frac{\Omega}{\Gamma_{\parallel}}\right)^2.
\end{split}
\label{eqb4}
\end{equation}
Note that $I_{coh}=\Gamma_{\parallel}S$ and $I_{incoh}=\Gamma_{\parallel}S^2$
at low driving fields ($S\ll 1$).

At low driving fields with $\Omega < |\Gamma_{\parallel}-2\gamma^{*}|/4$,
following the quantum regression theorem, we get the degree of first-order coherence
\begin{equation}
\begin{split}
\mathrm{g}^{(1)}(\tau)=&\exp{(-i\omega_{QD}\tau)}\times \Biggl\{\frac{\frac{\Gamma_{\parallel}}{2}}{\frac{\Gamma_{\parallel}}{2}+\gamma^{*}+\frac{\Omega^2}{\Gamma_{\parallel}}}
+\frac{1}{2}\exp{\left[-\left(\frac{\Gamma_{\parallel}}{2}+\gamma^{*}\right)\tau\right]}\\
&-\frac{\left(\frac{\Gamma_{\parallel}}{2}
-\gamma^{*}+2\lambda\right)^2}{8\lambda\left(\frac{3\Gamma_{\parallel}}{2}+\gamma^{*}-2\lambda\right)}
\exp{\left[-\left(\frac{3\Gamma_{\parallel}}{4}+\frac{\gamma^{*}}{2}-\lambda\right)\tau\right]}+(\lambda \rightarrow -\lambda)\Biggl\},
\end{split}
\label{eqb5}
\end{equation}
where $\lambda=\sqrt{(\Gamma_{\parallel}-2\gamma^{*})^2/16-\Omega^2}$ and the fourth term
in the bracket is the same as the third term with the sign of $\lambda$ reversed.
Compared with Eq. (\ref{eqb2}) at $\omega_L=\omega_{QD}$, we see
the first term is proportional to the intensity fraction of coherent component, so
this term stems from the coherent scattering with $\delta$-like laser spectrum.
The other three terms are proportional to the intensity fraction of incoherent components and
arise from the incoherent scattering with broad QD emission spectrum.

According to the Wiener-Khinchin theorem, the normalized emission spectrum can be
obtained from the Fourier transform of $\mathrm{g}^{(1)}(\tau)$, i.e.,
\begin{equation}
\begin{split}
S(\omega)=&\frac{1}{2\pi}\int_{-\infty}^{+\infty}d\tau \mathrm{g}^{(1)}(\tau)\exp{(i\omega\tau)}\\
=&\frac{\frac{\Gamma_{\parallel}}{2}}{\frac{\Gamma_{\parallel}}{2}+\gamma^{*}+\frac{\Omega^2}{\Gamma_{\parallel}}} \delta(\omega-\omega_{QD})+\frac{\left(\frac{\Gamma_{\parallel}}{2}+\gamma^{*}\right)/2\pi}{(\omega-\omega_{QD})^2+\left(\frac{\Gamma_{\parallel}}{2}+\gamma^{*}\right)^2}\\
&-\frac{1}{16\pi\lambda}\frac{\left(\frac{\Gamma_{\parallel}}{2}-\gamma^{*}+2\lambda\right)^2}{(\omega-\omega_{QD})^2+\left(\frac{3\Gamma_{\parallel}}{4}+\frac{\gamma^{*}}{2}-\lambda\right)^2}
+(\lambda \rightarrow -\lambda).
\end{split}
\label{eqb6}
\end{equation}

From the quantum regression theorem, we also get the degree of second-order coherence
\begin{equation}
g^{(2)}(\tau)=1-\left[\cosh{(\lambda \tau)}+\frac{\frac{3}{4}\Gamma_{\parallel}+\frac{\gamma^{*}}{2}}{\lambda}\sinh{(\lambda \tau)} \right]\exp {\left[-\left(\frac{3}{4}\Gamma_{\parallel}+\frac{\gamma^{*}}{2}\right)\tau\right]}.
\label{eqb7}
\end{equation}

Expressions for $g^{(1)}(\tau)$, emission spectra and $g^{(2)}(\tau)$ at higher laser powers with
$\Omega >|\Gamma_{\parallel}-2\gamma^{*}|/4$ can also be obtained in a similar way (not presented here).

However, this model is just an approximation to the QD-cavity system with the cavity field operator excluded.
This model cannot explain the laser-detuning and laser-power dependence of $\mathrm{g}^{(2)}(0)$
shown in Figs. 2(b) and 2(e) in the main text, therefore we use the master equation to simulate
the QD-cavity system to get exact results.

\section{EXPERIMENTS}

\subsection{Sample growth and device fabrication}
The sample wafer was grown on semi-insulating (100) GaAs substrates by solid-source molecular
beam epitaxy (Veeco Gen930). Low-density InAs QDs were realized by the in situ annealing
and the Indium gradient growth. The QDs were embedded in the center of a $\lambda$-thick GaAs
cavity ($\lambda=920\ $nm in this work) between the bottom 30 pairs and the top 18 pairs of
GaAs/Al$_{0.9}$Ga$_{0.1}$As DBRs. We employed phase-matching growth to precisely control the
thickness of each layer of DBR. First, we grew a reference sample with 6 pairs of upper DBR
and 8 pairs of lower DBR. Immediately after the growth, we measured the room-temperature
reflection spectrum and corrected the cavity mode shift of the reference sample
during the formal sample growth such that the cavity mode and quantum dot emission wavelengths
were perfectly matched to enhance the emission intensity of quantum dots.

The LCSPS device which contains a single In(Ga)As QD in circular micropillars is fabricated
from the sample wafer.
Before fabrication, we deposited a hard mask (a layer of 350-nm-thick $\mathrm{SiO_2}$) on the
as-prepared sample surface by plasma-enhanced chemical vapour deposition. Then
we created an array of micropillars with diameters between $2-3\ \mu$m on
the $\mathrm{SiO_2}$ mask by DWL66+ maskless laser lithography system.
After that micropillars were fabricated by a chlorine-based dry etching process
using an inductively coupled plasma (ICP) etching equipment (Plasma Pro 100 Cobra 300, Oxford Instruments).
Finally, the remaining $\mathrm{SiO_2}$ mask was removed by a fluorine-based
dry etching process.

We selected micropillars containing a single QD resonantly coupled to the fundamental cavity
mode with high Purcell factor. Then we chose the micropillars with close-to-zero cold-cavity reflectivity at the
center of the cavity mode. In this work, the demonstrated device has a diameter of $2.4\ \mu$m.

\subsection{Experimental setup for optical characterizations}
The schematic of the set-up for optical characterizations is presented in Fig. 7.
The sample was placed in a closed-cycle cryostat (attoDry 800) with a base temperature of $4.3-300\ $K.
To characterize the QD samples, a $5$-ps mode-locked Ti:sapphire laser with $80\ $MHz
repetition rate was used for non-resonant excitation and resonance fluorescence measurements as well as
time-resolved lifetime measurements. For bright-field coherent reflection spectroscopy measurement,
a laser light from a tunable cw Ti:sapphire oscillator (M SQUARED) with a linewidth $<100\ $kHz
(measured over a period of $100\ \mu$s) passes through a PBS and a $\lambda/4$ waveplate
and incident onto the sample through a $68 \times $ objective lens sitting in the cryostat.
The reflected light passes the $\lambda/4$ waveplate twice with its polarization
rotated by $90^{\circ}$ and then goes through the PBS without intensity loss.
This bright-field coherent reflection measurement can be easily switched to
cross-polarized resonance fluorescence setup by inserting two linear polarizers
before and after the PBS, which is a dark-field reflection measurement where most of
reflected light are blocked \cite{kuhlmann13b}.

The reflected light was collected by a single-mode fibre and then was guided to
different interferometers to characterize the device performance, including
Hanbury Brown-Twiss (HBT) interferometer to measure the single photon purity,
Michelson interferometer to measure the first-order coherence, HOM interferometer
to measure two-photon interference and photon indistinguishability, scanning
Fabry-P\'{e}rot interferometer with a resolution of $31\ $MHz and a free spectral
range of $20\ $GHz to measure the linewidth and Mollow triplet spectrum of QD
emission. A 4-channel superconducting nanowire single photon detector (SNSPD)
(Scontel) with time resolution $20\ $ps and efficiency $86\pm 1\%$ at $900\ $nm was used
for single-photon detection. An 8-channel time-correlated single photon counter (Swabian
Instruments) with time jitter $3\ $ps was used for correlation measurements and
time-resolved measurements. The overall collection efficiency from the first collection
lens sitting in the cryostat to the SNSPDs is around $25\%$.

\subsection{Measurement of QD radiative lifetime and Purcell factor}
When QD is tuned in resonance with the cavity mode, we measure the transient reflectivity with
a 5-ps mode-locked Ti:Saphire laser with its central wavelength at the QD transition.
A Lok-to-Clock Module (Model 3930) is used to stabilize the
repetition rate to $80\ $MHz and reduce the time jitter. Apart from the sharp
instrument response function (IRF), we observe a mono-exponential decay curve (Fig. 8, green line)
and extract the lifetime of QD exciton in resonance with the cavity mode as $\tau_{on}=57\ $ps.
This corresponds to $\Gamma_{\parallel}/2\pi = 2.8\ $GHz which is the cavity-enhanced QD spontaneous emission rate.

When QD is off resonance with the cavity mode by temperature tuning, we measure the QD
lifetime by collecting the resonance fluorescence intensity (Fig. 8, red line), and get $\tau_{off}=450\ $ps.
This corresponds to $\gamma_{\parallel}/2\pi = 0.35\ $GHz which is the QD spontaneous emission rate
into leaky modes. The observation of the beating signal with a period of $1100\ $ps
indicates our device contains a neutral QD with a fine-structure splitting of $0.91\ $GHz which is much smaller
than the QD linewidth $\Gamma_{\parallel}/2\pi = 2.8\ $GHz and can be neglected in our discussions.

The lifetime measurements yield a Purcell factor $F_p=\frac{\tau_{off}}{\tau_{on}}-1=6.9$.
From the measured coherent reflection spectra in Fig. 2(c), we obtain the quality factor $Q=8937$ of the
fundamental cavity mode, corresponding to $\kappa/2\pi=36.8\ $GHz. Assuming the QD is located at
the anti-node of cavity mode, we get the QD-cavity interaction strength $g/2\pi=4.7\ $GHz
based on the expression for the Purcell factor $F_p=4g^2/(\kappa \gamma_{\parallel})$.

\subsection{Observations of Mollow triplet and Rabi oscillations in reflected light}
As discussed in the main text, LCSPS works at lower
powers ($\Omega < \Gamma_{\parallel}/\sqrt{2}$) where the coherent reflection
dominates the incoherent reflection.

At higher powers with $\Omega > \Gamma_{\parallel}$, the QD gets saturated and the reflected light is dominated by
the QD emission (incoherent scattering). In frequency domain, we observe the Mollow triplet
in optical spectra of
reflected light [Fig. 9(a)] measured with a scanning Fabry-P\'{e}rot interferometer
(resolution $31\ $MHz, free spectral range $20\ $GHz).
In the time domain, Rabi oscillations are observed in the second-order correlation
function $\mathrm{g}^{(2)}(\tau)$ [Fig. 9(c)].  Note that $\mathrm{g}^{(2)}(0)$ gets worse
at high laser powers as predicted in Fig. 1(g).

Due to the Purcell effect, the cavity-enhanced QD linewidth is $F_p+1=7.9$ times
as broad as the QD linewidth without the cavity.
Rabi frequency exhibits a linear relationship with the square root of laser power [Fig. 9(b)].
To achieve the same Rabi frequency, the required laser power in our device is about
two orders of magnitude lower than a QD without cavity enhancement \cite{vamivakas09}.
This is due to the enhanced electric field strength in the cavity.

\newpage

\begingroup
%\begin{figure}[ht]
\centering
\includegraphics* [bb= 62 298 532 762, clip, width=12cm, height=12cm]{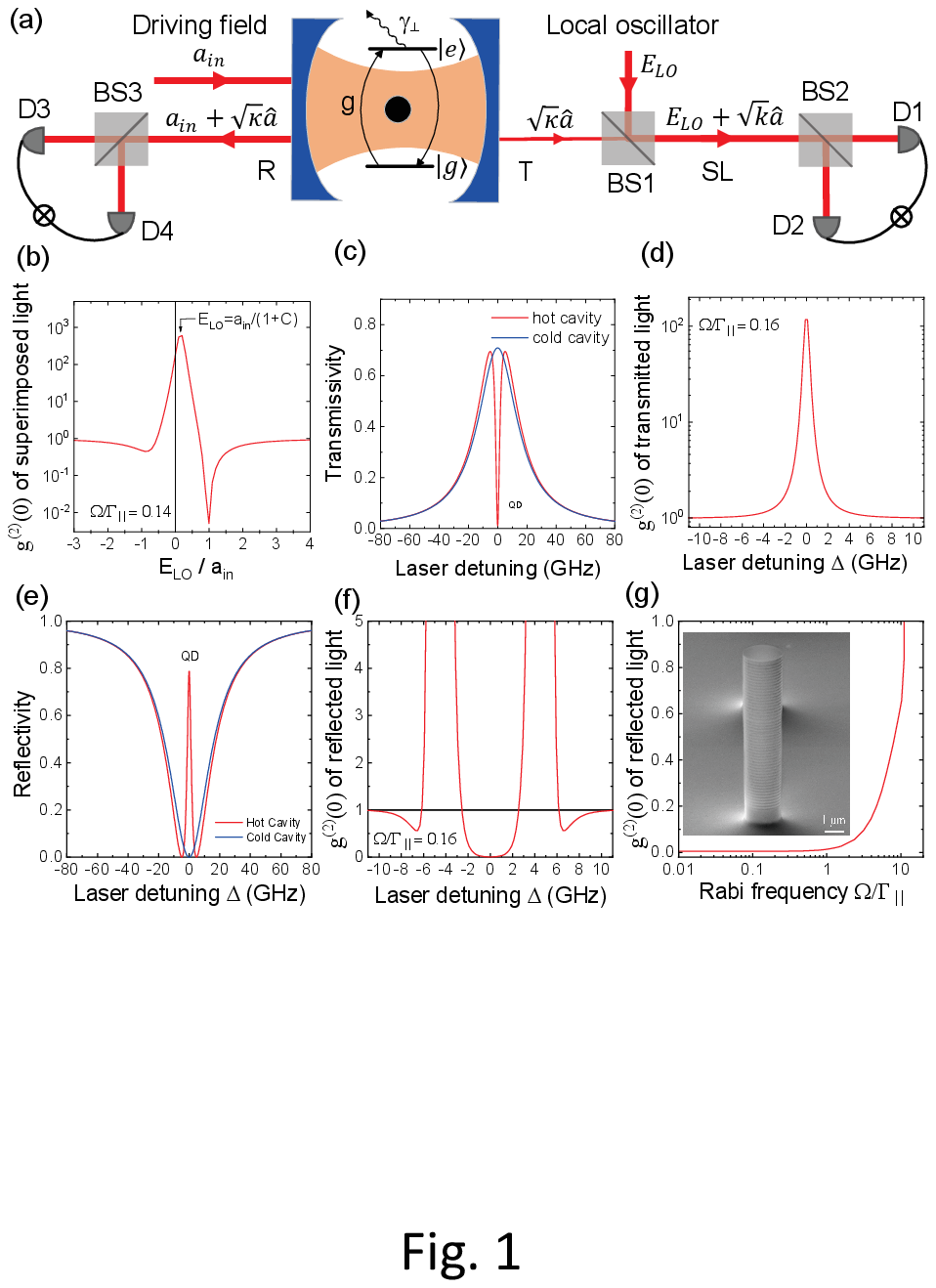}
\captionof{figure}{(color online) Working principle of the laser-converted single-photon source via interference.
(a) The device consists of a single two-level quantum emitter resonantly coupled to a symmetric double-sided
microcavity in the Purcell regime. The cavity is coherently driven by a laser field $a_{in}$.
The transmitted light is superimposed with a local laser field $E_{LO}$ on a beam splitter BS1.
The photon statistics of the superimposed light can be
tuned from anti-bunching to super-bunching by controlling the interference between the transmitted field and
local laser field. As a special case with $E_{LO}=a_{in}$, the reflected light consists of a superposition of the driving field
and the cavity output field (i.e., the transmitted field).
The destructive interference between the driving field and the
cavity output field results in the anti-bunching of reflected light.
BS1-BS3: 50:50 non-polarizing beam splitters, D1-D4: superconducting nanowire single photon detectors.
(b) Calculated $\mathrm{g}^{(2)}(0)$ of the superimposed light versus the relative amplitude of
the local laser field compared to the driving field.
(c) Calculated coherent transmission spectra of the hot cavity (red) and cold cavity (blue).
The cavity side leakage rate is taken to $0.08\kappa$.
(d) Calculated $\mathrm{g}^{(2)}(0)$ of transmitted light versus the frequency detuning of driving field.
(e) Calculated coherent reflection spectra of the hot cavity (red) and cold cavity (blue).
(f) Calculated  $\mathrm{g}^{(2)}(0)$ of reflected light versus the frequency detuning of driving field.
(g) Calculated $\mathrm{g}^{(2)}(0)$ of reflected light versus the driving strength at zero detuning.
Inset is a SEM image of our device consisting of a single QD coupling to a pillar microcavity.
The calculations in (b)-(g) are performed using the master equation or Heisenberg-Langevin equations of motions (see Appendix A).
We take $g/2\pi=4.7\ $GHz, $\kappa/2\pi=36.8\ $GHz and $\gamma_{\perp}/2\pi=0.18\ $GHz measured in our device (see Appendix C).}
\label{fig1}
%\end{figure}
\endgroup

\newpage

\begingroup
%\begin{figure}[ht]
\centering
\includegraphics* [bb= 115 495 480 699, clip, width=14cm, height=7.7cm]{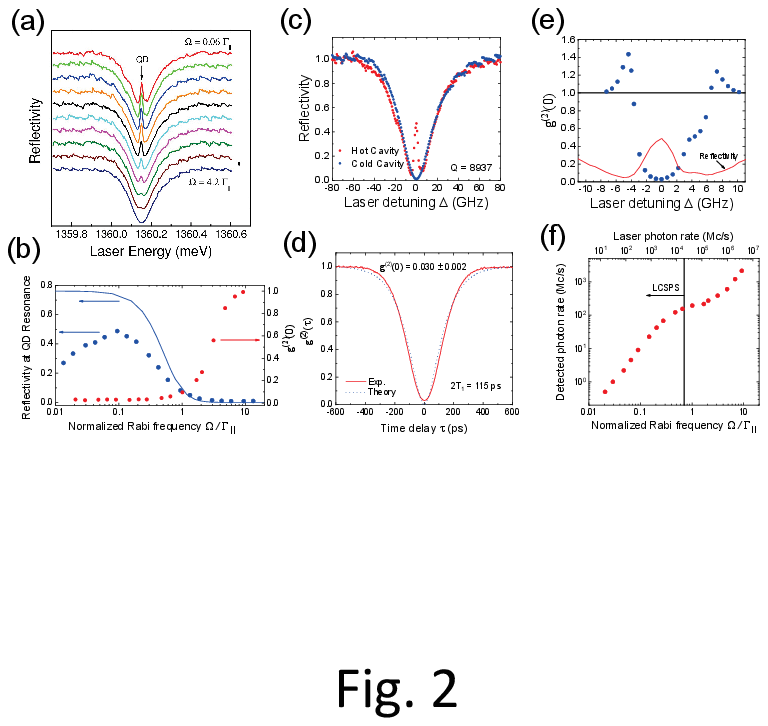}
\captionof{figure}{(color online) Implementation of the laser-converted single-photon source.
(a) Reflection spectra recorded by scanning the driving laser frequency at
different powers. A reflection peak due to the QD resonance is observed
inside the broad cavity resonance, but tends to wash out with increasing the driving laser power.
(b) Laser-power dependent coherent reflectivity at the QD resonance (blue, dots) and the
second-order autocorrelation $\mathrm{g}^{(2)}(0)$ (red, dots) of reflected light with
the laser frequency fixed at the QD resonance. The calculated coherent QD reflectivity (blue, solid line)
using the master equation is plotted here for comparison.
(c) A comparison of measured reflection spectra for the hot cavity (red) and cold cavity (blue).
(d) Measured second-order autocorrelation $\mathrm{g}^{(2)}(\tau)$ of
reflected light with the laser frequency fixed at the QD resonance. $\mathrm{g}^{(2)}(0)=0.030\pm 0.002$
is achieved. The dotted curve is the fitting results using either the master equation in Appendix A or
Eq. (\ref{eqb7}) in Appendix C.
(e) Measured second-order autocorrelation $\mathrm{g}^{(2)}(0)$ of reflected light
when the laser frequency is scanned cross the QD resonance. The driving laser power is fixed at
$\Omega = 0.11\Gamma_{\parallel}$. The measured reflection spectrum of hot cavity is a guide to the eye.
(f) The single-photon count rate detected with a SNSPD as a function of the driving laser power.
Calibrated neutral-density filters are used when the count rate goes beyond the detection limit of SNSPD.}
\label{fig2}
%\end{figure}
\endgroup

\newpage

\begingroup
%\begin{figure}[ht]
\centering
\includegraphics* [bb= 107 392 461 702, clip, width=12cm, height=10.5cm]{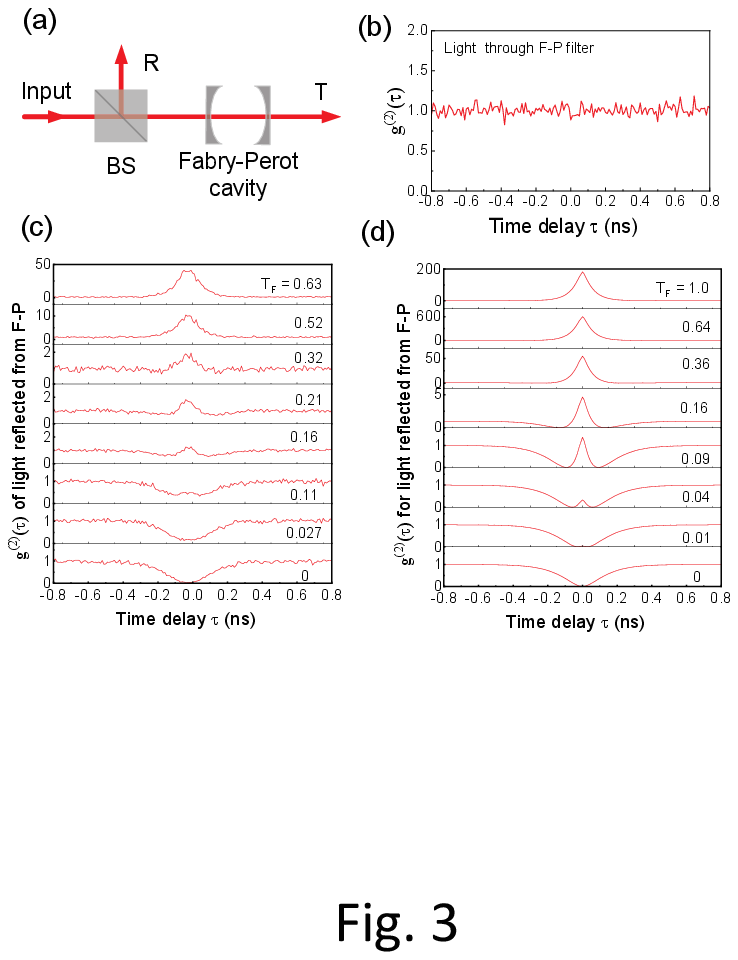}
\captionof{figure}{(color online) Photon statistics of coherent and incoherent components of LCSPS.
(a) Experimental setup to separate coherent and incoherent components using a Fabry-Perot
filter with a free spectral range of $20\ $GHz, a resolution of $31\ $MHz and a maximum transmission $T_F=63\%$.
The filter's resolution is much larger than the laser linewidth ($<100\ $kHz) but much smaller than the cavity-enhanced
QD linewidth ($\Gamma_{\parallel}=2.8\ $GHz) such that coherent component passes through the filter and
the incoherent light gets reflected from the filter. The relative amplitude of coherent and incoherent components
is controlled by slightly detuning the filter's modes. BS: non-polarizing beam splitter.
(b) Measured second-order autocorrelation $\mathrm{g}^{(2)}(\tau)$ of coherent component passing
through the Fabry-Perot filter. Photons in the coherent component are uncorrelated as expected.
(c) Measured and (d) Calculated second-order autocorrelation $\mathrm{g}^{(2)}(\tau)$ of the
remaining light reflected from the Fabry-Perot filter. The photon statistics changes from
anti-bunching to super-bunching with increasing the filter's transmission $T_F$.
The driving power is kept at $\Omega = 0.14\Gamma_{\parallel}$ for (b)-(d).}
\label{fig3}
%\end{figure}
\endgroup

\newpage

\begingroup
%\begin{figure}[ht]
\centering
\includegraphics* [bb= 88 330 510 529, clip, width=12cm, height=5.4cm]{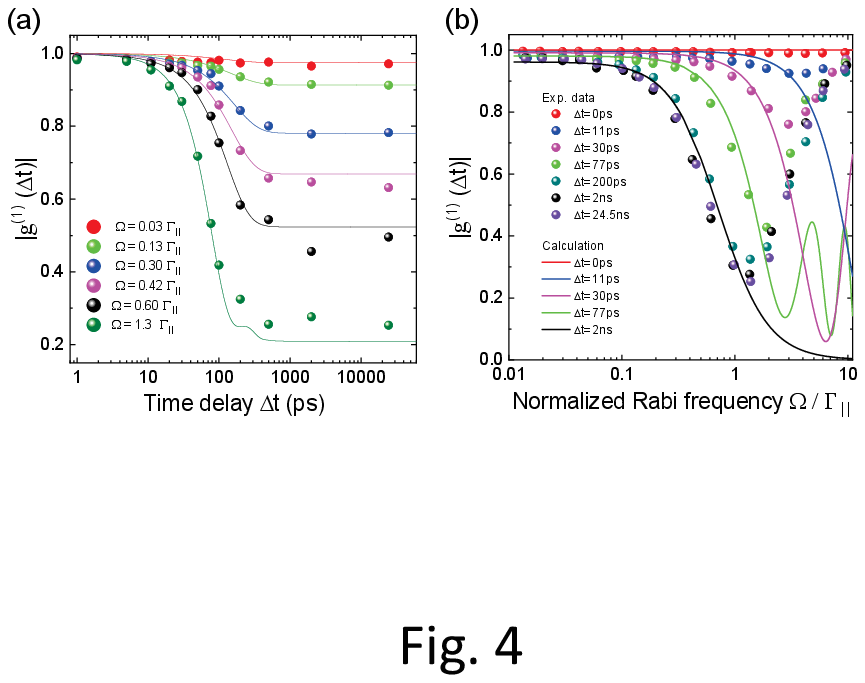}
\captionof{figure}{(color online) First-order coherence measured by a Michelson interferometer.
(a) Measured degree of first-order coherence $\mathrm{g}^{(1)}(\tau)$ of reflected light versus
time delay at different driving powers.
(b) Measured degree of first-order coherence $\mathrm{g}^{(1)}(\tau)$ of reflected light as a function
of the driving power at different time delays. The solid curves in (a)  and (b)
are calculated using the master equation in Appendix A and  Eq. (\ref{eqb5}) in Appendix B which
give the same results.}
\label{fig4}
%\end{figure}
\endgroup

\newpage

\begingroup
%\begin{figure}[ht]
\centering
\includegraphics* [bb= 90 468 514 684, clip, width=14cm, height=7cm]{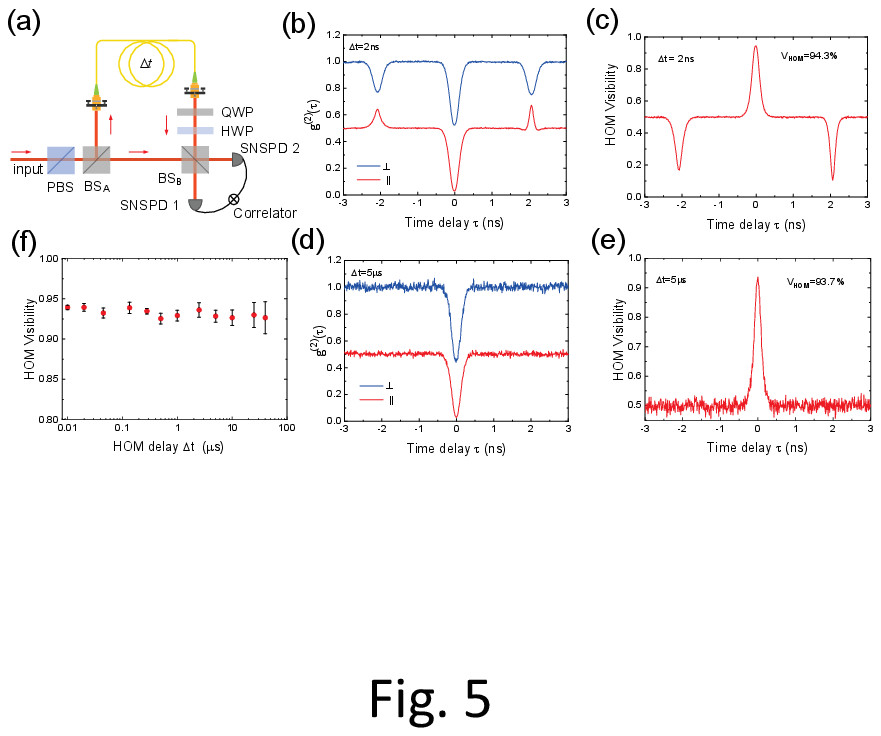}
\captionof{figure}{(color online) Two-photon interference and photon indistinguishability.
(a) The AMZI setup to measure TPI (or HOM) interference.
Either free-space or fibre delays are used to vary $\Delta t$. BS$_\mathrm{A}$ and BS$_\mathrm{B}$: beam splitters,
PBS: polarizing beam splitter, SNSPD 1 and SNSPD 2: superconducting nanowire single photon detectors, QWP: quarter-wave plate, HWP: half-wave plate.
(b) Two-photon interference measurements for cross polarization (blue)
and parallel polarization (red) configuration at $\Delta t=2$ ns.
(c) Two-photon interference visibility versus time delay $\tau$ at $\Delta t=2$ ns. $V_{HOM}(0)=94.3\%\pm 0.2\%$.
(d) Two-photon interference measurements for cross polarization (blue)
and parallel polarization (red) configuration versus time delay $\tau$ at $\Delta t=5\ \mu$s using a $1$-km fibre delay.
(e) Two-photon interference visibility versus the time delay $\tau$ at $\Delta t=5\ \mu$s. $V_{HOM}(0)=93.7\%\pm 0.2\%$.
(f) Two-photon interference visibility $V_{HOM}(0)$ as a function of $\Delta t$.
All data points presented in (b)-(f) are raw data without
correcting experimental imperfections. The driving power is fixed at $\Omega = 0.13\Gamma_{\parallel}$.}
\label{fig5}
%\end{figure}
\endgroup

\newpage

\begingroup
%\begin{figure}[ht]
\centering
\includegraphics* [bb= 71 279 532 473, clip, width=12cm, height=4.9cm]{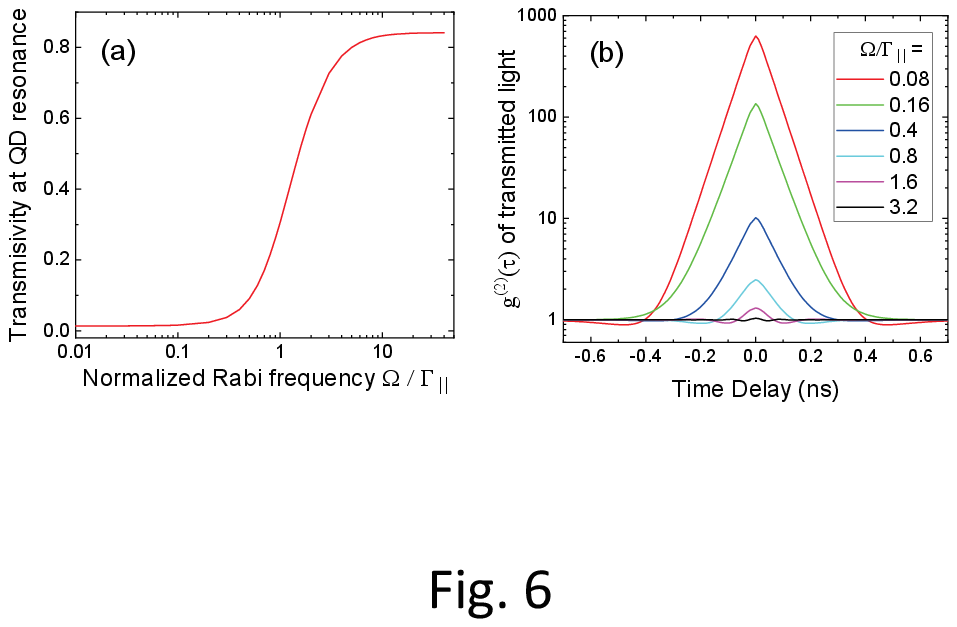}
\captionof{figure}{(a) Calculated coherent transmissivity at QD resonance versus
the driving powers using the master equation.
(b) Calculated second-order correlation $\mathrm{g}^{(2)}(\tau)$ of transmitted light at different
driving powers. In contrast to reflected light, transmitted light exhibits super-bunching.
Rabi oscillations appear at higher driving powers.}
\label{fig6}
%\end{figure}
\endgroup

\newpage

\begingroup
%\begin{figure}[ht]
\centering
\includegraphics* [bb= 14 199 580 641, clip, width=15cm, height=11.6cm]{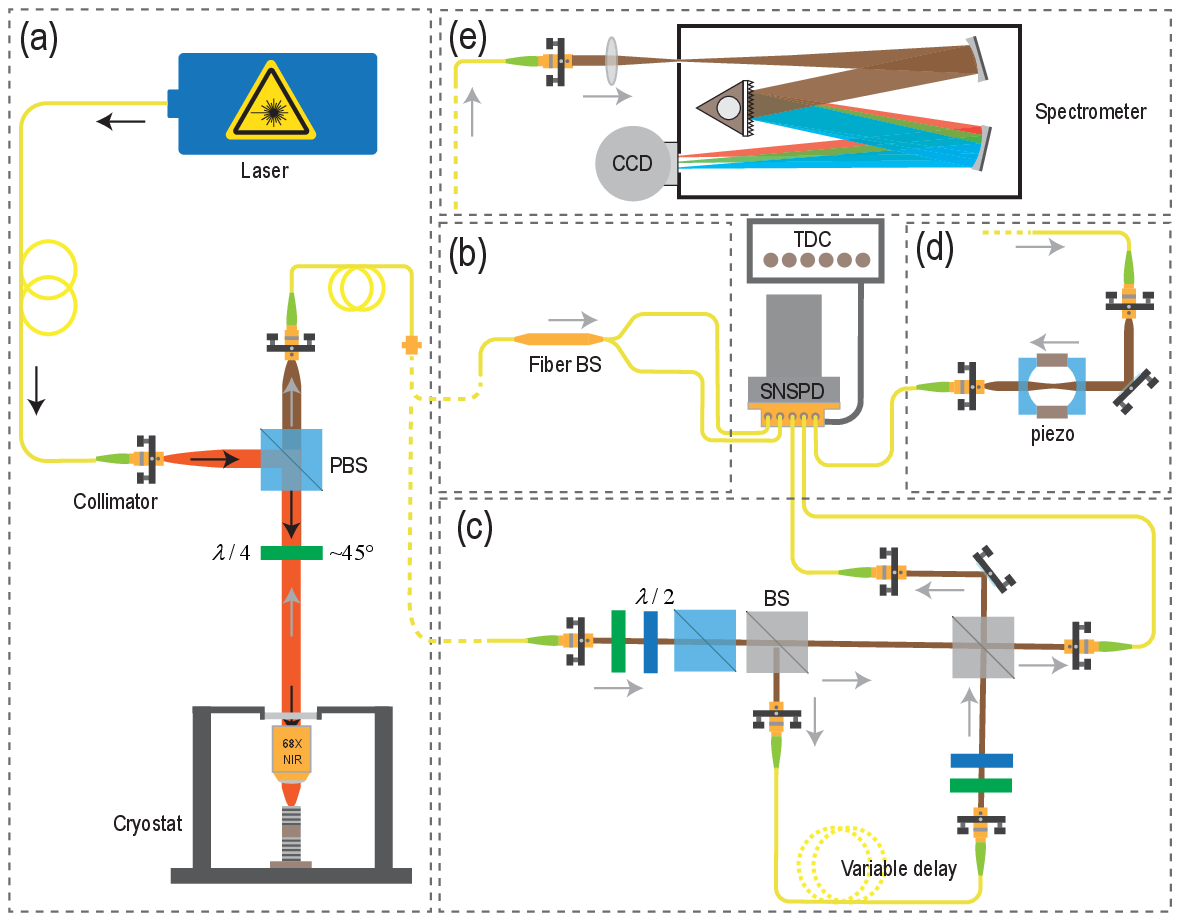}
\captionof{figure}{Schematic of setup for optical characterization of devices.
The setup consisting of five functional sections including: (a) cryogenic confocal
system for coherent reflection spectroscopy and cross-polarized resonance fluorescence
measurements. (b) Hanbury Brown - Twiss (HBT) measurement kit. (c) HOM/AMZ/Michelson interferometers.
(d) scanning Fabry-P\'{e}rot interferometer. (e) optical spectrometer. BS: beam splitter,
PBS: polarizing beam splitter, SNSPDs: multi-channel superconducting nanowire single photon detectors,
TDC: time-to-digital converter.}
\label{fig7}
%\end{figure}
\endgroup

\newpage

\begingroup

%\begin{figure}[ht]
\centering
\includegraphics* [bb= 175 388 439 582, clip, width=8cm, height=6cm]{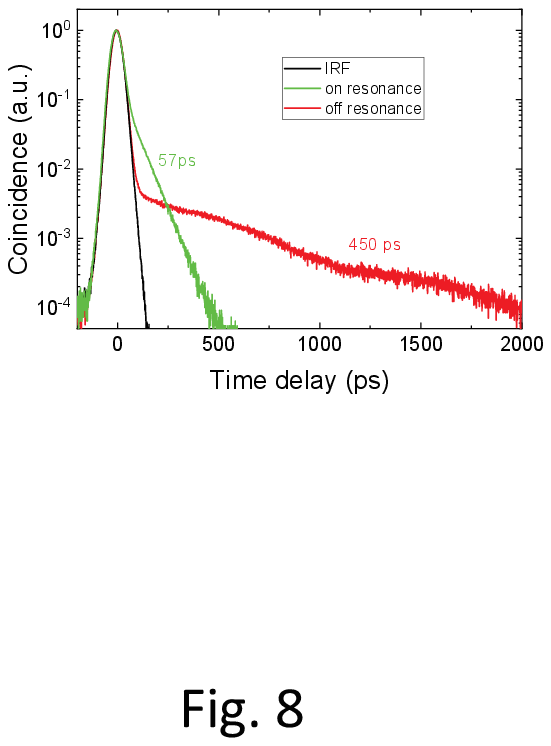}
\captionof{figure}{QD lifetime measurements.
The time-resolved measurement for QD in resonance with the cavity mode was done by monitoring the reflected light intensity,
and the time-resolved measurement for QD far-detuned from the cavity mode by temperature
tuning was performed by monitoring the resonance
fluorescence intensity under resonant excitation condition.}
\label{fig8}
%\end{figure}
\endgroup

\newpage

\begingroup

%\begin{figure}[ht]
\centering
\includegraphics* [bb= 65 386 528 572, clip, width=14cm, height=5.6cm]{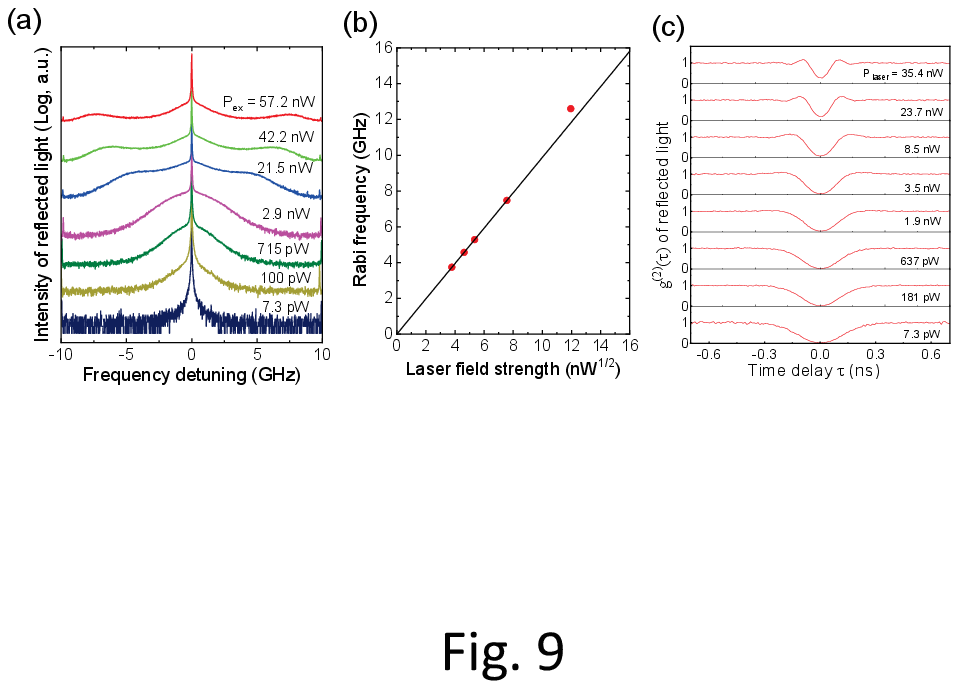}
\captionof{figure}{(a) Evolution of the Mollow triplet spectrum of reflected light as the incident laser power is increased. (b) Extracted Rabi frequency as a function of the square root of incident laser power on a linear scale. The data points exhibit a linear relationship between the Rabi frequency and the laser field strength. (c) Raw second-order autocorrelation data taken on the reflected light at different laser powers.}
\label{fig9}
%\end{figure}
\endgroup

\end{document}